\documentclass[%
 reprint,
 amsmath,amssymb,
]{revtex4-2}

\usepackage{graphicx}
\usepackage{dcolumn}
\usepackage{bm}
\usepackage{ulem}

\usepackage{amssymb} 
\usepackage{multirow}
\usepackage{booktabs}
\usepackage{color}     
\usepackage{amsmath}
\usepackage{multirow}  

\usepackage{hyperref} 
\hypersetup{
            colorlinks=true,
            linkcolor=blue,
            filecolor=magenta,      
            urlcolor=cyan, 
            citecolor=red,}
            
\begin{document}

\preprint{APS/123-QED}

\title{Many-body correlations as the origin of Gamow-Teller quenching in nuclear $\beta$-decay } 


\author{Hao Zhou}%
\affiliation{School of Physical Science and Technology, Southwest University, Chongqing 400715, China}%

\author{Long-Jun Wang}
\email{longjun@swu.edu.cn}
\affiliation{School of Physical Science and Technology, Southwest University, Chongqing 400715, China} 

\author{Yang Sun}
\affiliation{School of Physics and Astronomy, Shanghai Jiao Tong University, Shanghai 200240, China}%

\date{\today}

\begin{abstract}
The longstanding quenching problem of Gamow–Teller (GT) strength in nuclear $\beta$-decay is attributed to missing contributions in the transition operator and/or incomplete nuclear correlations in the many-body wavefunction. Recent studies have predominantly emphasized operator renormalization, including chiral two-body currents, while the effects of many-body correlations—especially in heavy open-shell nuclei—remain underappreciated. We present a large-configuration shell-model calculation that incorporates chiral two-body weak current and treats both mechanisms on equal footing. Taking the neutrinoless double $\beta$-decay candidate $^{76}$Ge as an example, we demonstrate that strong nuclear correlations drive a substantial portion of GT strength to high excitation energies, leading to a pronounced suppression of low-energy strength responsible for the apparent quenching. We identify that the quenching originates mainly from deformation, cross-shell correlations, and mixing among densely-spaced highly excited states. In contrast, the chiral two-body current contributes only a modest $5–15\%$ reduction, depending on the coupling constants employed.
Our results thus suggest many-body correlations as the primary origin of GT quenching and provide a unified microscopic explanation for this phenomenon in nuclear $\beta$-decay.

\end{abstract}

\maketitle



The quenching puzzle of Gamow–Teller (GT) strength in nuclear $\beta$-decay remains one of the central unresolved problems in nuclear structure and weak-interaction physics. It has long been known that theoretical predictions by the model-independent sum rule \cite{Lipparini_1989_Phys_Rep} appears to exceed systematically the GT strength observed in experiments \cite{Bertsch_1987_RPP, Osterfeld_1992_RMP}.
In the past decades, numerous studies have been dedicated to finding possible origins of this discrepancy. Research in nuclear physics has focused on the reduced nuclear matrix element of GT transition
\begin{eqnarray} \label{eq.def}
 \left\langle \Psi_{J_f^{\pi_f}}  \left\| \hat{ \mathcal{O}}^\lambda \right\| \Psi_{J_i^{\pi_i}} \right\rangle ,
\end{eqnarray}
where $\hat{ \mathcal{O}}$ is the GT operator of rank $\lambda$ and $\Psi$ represents many-body wave functions with definite angular momentum and parity $J^\pi$. To fit experimental data, spherical shell-model calculations within the full $0\hbar\omega$ model space have to add a quenching factor, $q\approx 0.75$, to Eq. (\ref{eq.def}) without knowing its exact sources. This is referred to as the {\it quenching puzzle} \cite{Brown_1985_quenching, Towner_1987_Phys_Rep, Martinez_Pinedo_1996_quenching} in $\beta$ decay. 

From Eq. (\ref{eq.def}), it is evident that there are only two possible sources to explain the puzzle: the renormalization of the GT
transition operator and the missing correlations in many-body wave functions. The study in Ref. \cite{Martinez_Pinedo_1996_quenching} found that different quenching factors are required in different mass regions and models, implying that the quenching has a structure-dependent nature.

In recent years, with respect to the first source, the role of the chiral two-body current (TBC)  has been extensively studied \cite{Javier2011PRL, Currents_PRL_2014, LJWang_current_2018_Rapid, Gysbers_2019_Nat_Phys, Jon_Engel_currents_PRC_2022}. Based on {\it ab initio} calculations, it was demonstrated that the TBC provide a small quenching with $q^2 \approx 0.9$ or even an enhancement for some light nuclei with $A \lesssim 45$, and a large quenching with $q^2 \approx 0.7$ for some doubly magic heavier nuclei \cite{Currents_PRL_2014, Gysbers_2019_Nat_Phys}. It was claimed that the discrepancy between the experimental and theoretical $\beta$ decay rates has been resolved \cite{Gysbers_2019_Nat_Phys}.

However, the second source, namely potentially missing correlations, which can lead to large uncertainties in many-body wave functions, remains unclear. Studies on heavier, open-shell nuclei are even more scarce. These nuclei are generally deformed, and their level density increases exponentially with excitation energy \cite{Guttormsen_Level_Density_EpJA_2015, JQWang_PRC_2023}; therefore, mixing with highly excited states can significantly affect the global GT strength distribution. Furthermore, the effect of TBC on GT transitions in heavier and open-shell nuclei may differ from that in light or doubly magic nuclei, and should be thoroughly discussed when addressing the quenching puzzle \cite{Jon_Engel_currents_PRC_2022}. 

In this Letter, we investigate the quenching problem from {\it both} sources and provide a {\it quantitative} analysis of the role of each source. Our calculations are based on a novel shell-model approach that combines the projection technique with a large configuration space while taking into account the chiral TBC exactly. We study several structural effects that had not been previously considered, which could strongly influence the {\it global} GT strength distribution through correlations across multiple major shells as well as configuration mixing with densely-spaced high-excition energy levels.  

In order to carry out this study, it is necessary to have a theoretical model that can realistically describe highly excited states of deformed heavy nuclei and incorporate chiral TBC in the calculation, which poses a great challenge to  traditional shell models and modern {\it ab initio} methods. Clearly, novel algorithm for construction of many-body configurations \cite{Sun_2016_Phys_Scr} and breakthroughs in computation \cite{Robledo_2009_PRC_R} are required. We compute the nuclear matrix elements in Eq. (\ref{eq.def}) by the recent extension of the projected shell model (PSM) \cite{LJWang_2014_PRC_Rapid, LJWang_2016_PRC}, in which many-body wavefunctions are constructed by a superposition of (angular-momentum) projected multi-quasiparticle (qp) configurations. In the PSM, the wavefunction corresponding to $n^{\rm th}$ eigenstate of definite spin $J$ is expressed as 
\begin{eqnarray} \label{eq.wave_function}
  | \Psi^{n}_{JM} \rangle = \sum_{K\kappa} f_{K\kappa}^{Jn} \hat{P}_{MK}^{J} | \Phi_{\kappa} \rangle ,
\end{eqnarray}
in which $f$ labels the  expansion coefficients in the projected basis and $\hat{P}_{MK}^{J}$ the angular-momentum-projection operator
\begin{eqnarray} \label{AMP_operator}
    \hat{P}^{J}_{MK} = \frac{2J + 1}{8\pi^2} \int d\Omega D^{J\ast}_{MK} (\Omega) \hat{R} (\Omega) .
\end{eqnarray}
$\hat{R}$ and $D_{MK}^{J}$ \cite{varshalovich1988quantum}  in (\ref{AMP_operator}) are the rotation operator and Wigner $D$-function \cite{BLWang_2022_PRC}, respectively, with $K$ ($M$) being projection of angular momentum in the intrinsic (laboratory) frame. The multi-qp configurations, $|\Phi_{\kappa}\rangle$ in Eq. (\ref{eq.wave_function}), are explicitly given for even-even (ee) and odd-odd (oo) nuclei as \cite{BLWang_1stF_2024, LJWang_2014_PRC_Rapid} 
\begin{align} \label{eq.config}
  \textrm{ee}: \big\{ & |\Phi \rangle, 
               \hat{a}^\dag_{n_i} \hat{a}^\dag_{n_j} |\Phi \rangle,
               \hat{a}^\dag_{p_i} \hat{a}^\dag_{p_j} |\Phi \rangle,
               \hat{a}^\dag_{n_i} \hat{a}^\dag_{n_j} \hat{a}^\dag_{p_k} \hat{a}^\dag_{p_l} |\Phi \rangle, \nonumber\\ 
             & \hat{a}^\dag_{n_i} \hat{a}^\dag_{n_j} \hat{a}^\dag_{n_k} \hat{a}^\dag_{n_l} |\Phi \rangle, 
               \hat{a}^\dag_{p_i} \hat{a}^\dag_{p_j} \hat{a}^\dag_{p_k} \hat{a}^\dag_{p_l} |\Phi \rangle   \big\}, \nonumber \\
  \textrm{oo}: \big\{ & \hat{a}^\dag_{n_i} \hat{a}^\dag_{p_j}|\Phi \rangle, 
               \hat{a}^\dag_{n_i} \hat{a}^\dag_{n_j} \hat{a}^\dag_{n_k} \hat{a}^\dag_{p_l} |\Phi \rangle,
               \hat{a}^\dag_{n_i} \hat{a}^\dag_{p_j} \hat{a}^\dag_{p_k} \hat{a}^\dag_{p_l} |\Phi \rangle \big\}, 
\end{align}
with $|\Phi \rangle$ being deformed qp vacuum and $\hat{a}^\dag_n  (\hat{a}^\dag_p)$ the neutron (proton) qp creation operator associated with $|\Phi \rangle$. The coefficients $f$ in (\ref{eq.wave_function}) can be obtained by solving the Hill-Wheeler equation with a chosen two-body Hamiltonian \cite{PSM_review}. 

It is a nontrivial task to work simultaneously with the large configuration space in (\ref{eq.config}) and the TBC for arbitrarily heavy and deformed nuclei. Therefore, we adopt a simplified form of Hamiltonian. In the present calculation, we use an effective Hamiltonian with separable forces, including the monopole and quadrupole terms in both the particle-hole and particle-particle channels plus the two-body GT force \cite{BLWang_1stF_2024, LJWang_2018_PRC_GT}. According to Dufour and Zuker \cite{Dufour_1996_PRC}, these terms are the leading components of any realistic forces, which must be present to fulfill the basic requirement for structure calculations. 
With the long-wavelength approximation, the transition operator in nuclear weak (single $\beta$) decays corresponds to the single-nucleon weak current which includes the vector and axial-vector pieces. When considering only the axial current in the limit of zero momentum transfer, the one-body current (OBC) operator is written in the first-quantization form as \cite{LJWang_current_2018_Rapid, Jon_Engel_currents_PRC_2022, Park_2003_PRC},
\begin{eqnarray} \label{eq.1body_current}
  \hat{\bm{\mathcal J}}_{1b}(\bm x) = - g_A \sum_{i=1}^A \bm\sigma_i \tau^-_i \delta(\bm x - \bm r_i), 
\end{eqnarray}
where $g_A \approx 1.27$ is the weak axial-vector coupling constant \cite{Axial_Vector_PRL_2019}, $\bm\sigma$ the Pauli spin operator, and $\tau^-$ the isospin lowering operator with the convention $\tau^-|n\rangle = |p\rangle$. 

To further consider the chiral TBC, we follow \cite{Park_2003_PRC} by neglecting the term with the coefficient $c_6$ and the two-body pion pole terms \cite{Hoferichter_PLB_2015}. An additional factor of $-\frac{1}{4}$ is introduced to the contact term \cite{LJWang_current_2018_Rapid, Gysbers_2019_Nat_Phys, Krebs_AP_2017}. The leading space piece of the axial two-body current operator in coordinate space reads \cite{LJWang_current_2018_Rapid, Jon_Engel_currents_PRC_2022}
\begin{widetext}
\begin{eqnarray} \label{eq.2body_current}
  \hat{\bm{\mathcal J}}_{2b}(\bm x) &=& \sum_{k<l}^A \bm J_{kl}(\bm x), \nonumber \\ 
  \bm J_{kl}(\bm x) &=& \frac{2 \bar c_3 g_A}{m_N F^2_\pi} \left\{   m^2_\pi\left[\left( \frac{\bm\sigma_l}{3} - \bm\sigma_l \cdot \hat{\bm r} \hat{\bm r} \right) Y_2(r) - \frac{\bm\sigma_l}{3} Y_0(r) \right] + \frac{\bm\sigma_l}{3} \delta(\bm r)  \right\} \tau^-_l \delta(\bm x - \bm r_k) + (k \leftrightarrow l) \nonumber \\
  && + \left( \bar c_4 + \frac{1}{4}\right) \frac{g_A}{2 m_N F_\pi^2} \left\{ m_\pi^2 \left[ \left( \frac{\bm\sigma_\times}{3} - \bm\sigma_k \times \hat{\bm r} \bm\sigma_l \cdot \hat{\bm r} \right) Y_2(r) - \frac{\bm\sigma_\times}{3} Y_0(r) \right] + \frac{\bm\sigma_\times}{3} \delta(\bm r) \right\} \tau^-_\times \delta(\bm x - \bm r_k) + (k \leftrightarrow l) \nonumber \\
  && - \frac{g_A}{4m_N F_\pi^2} \left[ 2\bar d_1 \left( \bm \sigma_k \tau^-_k + \bm\sigma_l \tau^-_l\right) + \bar d_2 \bm\sigma_\times \tau^-_\times \right] \delta(\bm r) \delta (\bm x - \bm r_k),
\end{eqnarray}
\end{widetext}
where $F_\pi = 92.4$ MeV is the pion decay constant, $m_\pi$ is the pion mass, $\bm r = \bm r_k - \bm r_l$, and $\hat{\bm r} \equiv \frac{\bm r}{r}$. The Yukawa functions are $Y_0(r) = \frac{e^{-m_\pi r}}{4\pi r}$ and $Y_2(r) = \frac{1}{m_\pi^2} r \frac{\partial}{\partial r} \frac{1}{r} \frac{\partial}{\partial r} Y_0(r)$. The compound spin and isospin operators are $\bm\sigma_\times = \bm\sigma_k \times \bm\sigma_l$ and $\tau^-_\times = (\tau_k \times \tau_l)^-$ \cite{Park_2003_PRC}. The dimensionless low-energy constants (LECs) are $\bar c_3, \bar c_4, \bar d_1, \bar d_2$ with the definition $\bar c_{3, 4} = m_N c_{3, 4}$, $\bar d_{1, 2} = \frac{m_N F_\pi^2}{g_A} d_{1, 2}$ \cite{Park_2003_PRC} and $\bar c_{D} \equiv \bar d_1 + 2 \bar d_2$. 

In the present calculation, all the model parameters are the same as in Ref. \cite{LJWang_2018_PRC_GT}. To consider the contributions of the OBC in (\ref{eq.1body_current}), we adopt the reduced one-body transition density by the Pfaffian algorithm as in Refs. \cite{FGao_PRC2023, BLWang_1stF_2024}. To further account for contributions of the TBC in (\ref{eq.2body_current}), we have, for the first time in this work, accomplished derivations for the reduced two-body transition density within the projection theory (detailed derivations are provided in Supplemental Material \cite{Suppl_Material}).

\begin{figure}
\begin{center}
  \includegraphics[width=0.48\textwidth]{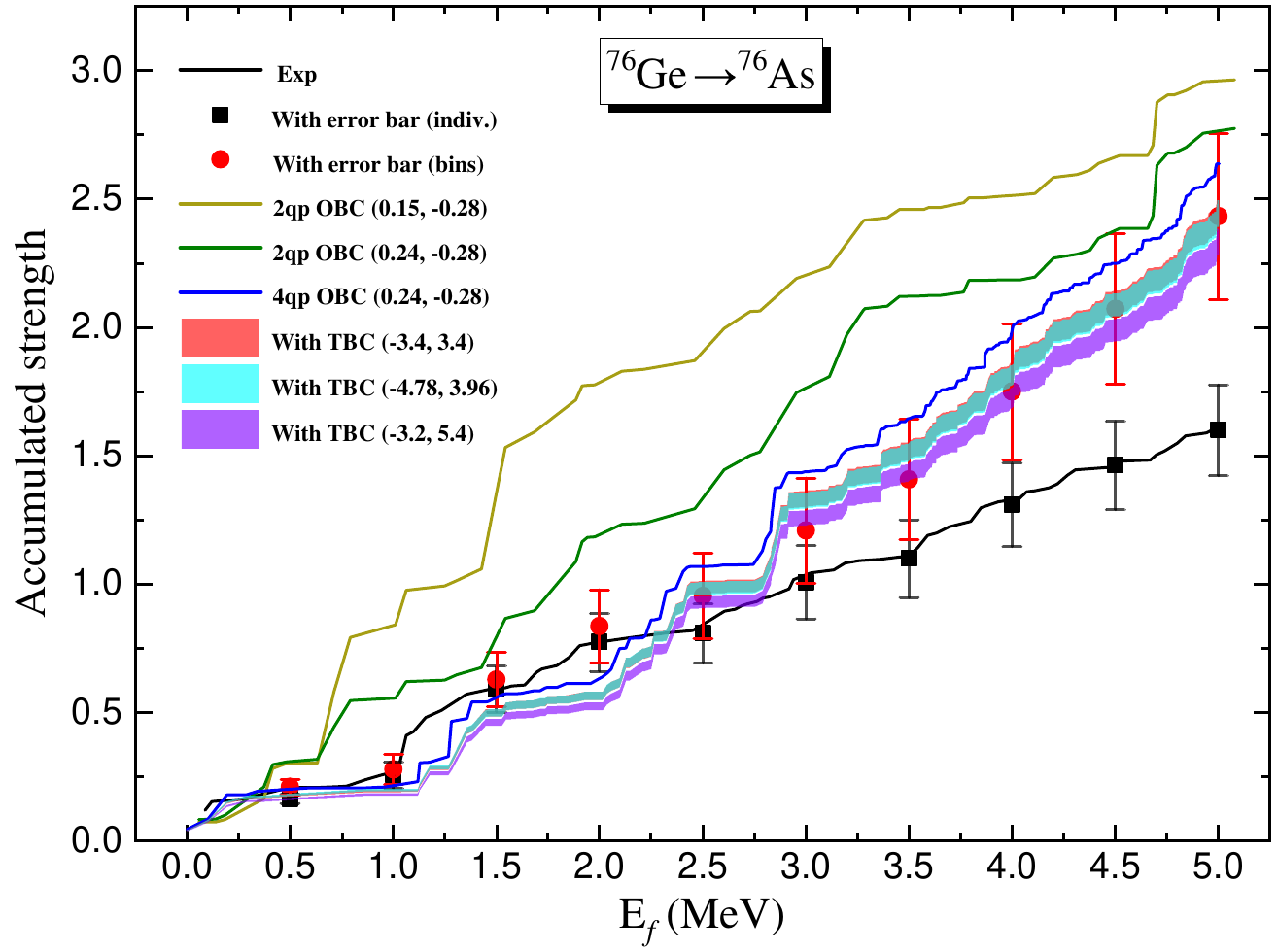}
  \caption{\label{fig:sum_BGT} (Color online) Cumulative sum of GT transition strengths from the $0_1^+$ ground state of $^{76}$Ge to all $1_f^+$ states of $^{76}$As within the excitation $E_f \leqslant 5.0$ MeV. The charge-exchange reaction data  \cite{76Ge_BGT_PRC_2012} are compared to the PSM results with different calculation conditions: (1) with different deformation parameters, (2) with small (up to 2qp) and large (up to 4qp) configuration spaces, and (3) for cases when only one-body current (OBC) or when both OBC and two-body current (TBC) are considered. See the text for details.  } 
\end{center}
\end{figure}

We take the neutrinoless double $\beta$-decay candidate $^{76}$Ge as example for discussion. In Fig. \ref{fig:sum_BGT}, the charge-exchange reaction data \cite{76Ge_BGT_PRC_2012}, with (red circle) and without (black square) GT resonance contributions, are shown by cumulative strength sum of GT transitions. All the theoretical results are obtained with a fixed $g_A = 1.0$. As the discussion extends to high excitations across the resonance region, our results beyond 2.0 MeV are compared to the red-circle data. We discuss sequentially effect of deformation and configuration mixing, and that of the chiral TBC within the large multi-qp configurations.

We first discuss the dependence on deformation parameter that defines the mean field, while other calculation conditions are fixed with (1) inclusion of only OBC and (2) a smaller configuration space with 2-qp states. For the daughter nucleus $^{76}$As, we adopt quadrupole deformation parameter $\varepsilon_2 = -0.28$ as M\"oller {\it et al.} \cite{moller2016} suggest that it is an oblately deformed nucleus. With the suggested value $\varepsilon_2 = 0.15$ for the parent nucleus $^{76}$Ge \cite{moller2016, P_Sarriguren_PRC_2012}, the obtained result (yellow line in Fig. \ref{fig:sum_BGT}) overestimates the data by more than a factor of two on average. When a larger value $\varepsilon_2 = 0.24$ deduced from experimental $E2$ transition probability \cite{Raman_ADNDT_2001} is used, the calculated result (green line in Fig. \ref{fig:sum_BGT}) is considerately suppressed. We can understand the result as follows. Since the intrinsic deformations of the daughter and parent nuclei are now similar in size ($|-0.28|$ versus 0.24) but opposite in sign (oblate versus prolate), their structural differences reach their maximum, thereby largely reducing the overlap between their wave functions and thus reducing the nuclear matrix element in (\ref{eq.def}) \cite{Lv_PRC_2022}.

\begin{figure}
\begin{center}
  \includegraphics[width=0.48\textwidth]{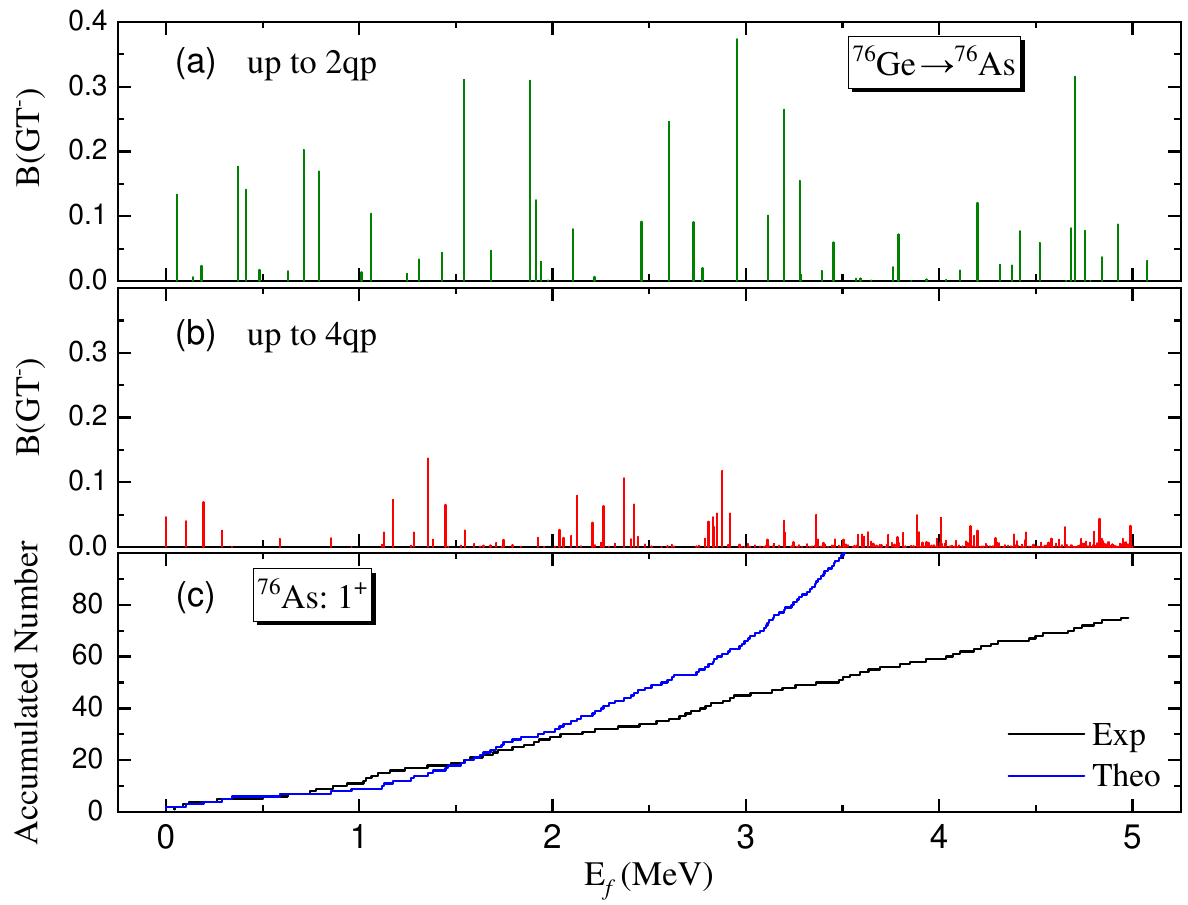}
  \caption{\label{fig:each_BGT} (Color online) Calculated GT strength distribution for individual transitions (with only one-body current considered), for (a) small (up to 2qp) configuration space, and (b) large (up to 4qp) configuration space. (c) Calculated cumulative sum of numbers of $1^+$ levels in $^{76}$As for case (b), as compared with the known experimental $1^+$ levels in Ref. \cite{NNDC}. } 
\end{center}
\end{figure}

The above calculation considered only 2-qp states in the configuration space, corresponding to configurations of one-pair breaking. Going up to higher excitations, configurations of simultaneous breaking of two nucleon-pairs (4-qps) dominate the structure  \cite{JQWang_PRC_2023}. As shown in Fig. \ref{fig:sum_BGT}, with inclusion of 4-qp configurations in (\ref{eq.config}), the calculated GT strength (blue line) is overall suppressed further, thus correctly describing the data for $E_f < 2.5$. 


It is important to recognize that the suppressed strengths in the low-energy region do not disappear, but shift to higher-energy region. In Fig. \ref{fig:each_BGT}, we compare GT strength distribution calculated with smaller (up to 2-qp) and larger (up to 4-qp) configuration spaces. It is seen from Fig. \ref{fig:each_BGT}(b), while strengths of \ref{fig:each_BGT}(a) are overall suppressed, many 4-qp states appear at high energies for $E_f > 3.0$ MeV with small strengths. A quantitative counting indicates that there are about 150 $1^+$ levels in the $E_f \approx 4.0-5.0$ MeV bin of Fig. \ref{fig:each_BGT}(b), instead of 15 $1^+$ levels in the same bin of Fig. \ref{fig:each_BGT}(a). In Fig. \ref{fig:each_BGT}(c), we give the counting result for the $1^+$ levels in $^{76}$As when 4-qp configurations are included. Our result coincides with experimental data up to $E_f=2$ MeV, beyond which our number rises exponentially while the data are not due to the incomplete measurement.

\begin{figure*}
\begin{center}
  \includegraphics[width=1.00\textwidth]{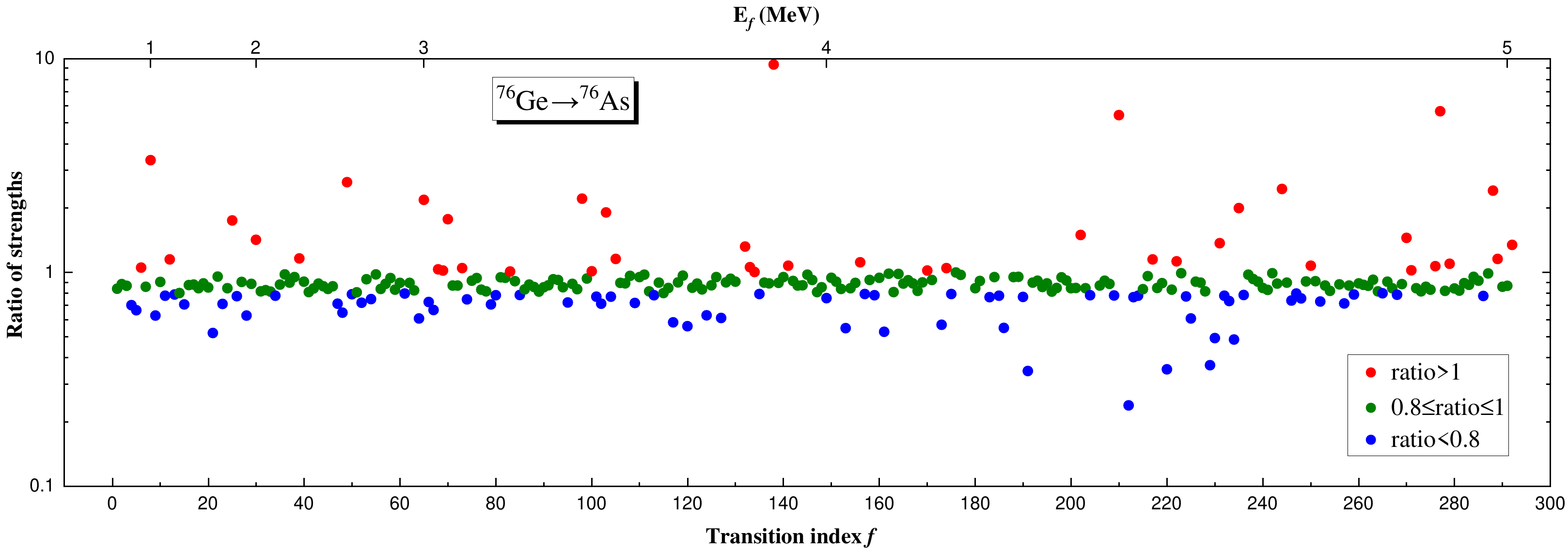}
  \caption{\label{fig:quenching} (Color online) The ratio of strengths calculated with the two-body currents to those without the two-body currents, i.e., $  \langle \Psi_{1^+_f}  \| (\hat{ \mathcal{J}}_{1b} + \hat{ \mathcal{J}}_{2b}) \| \Psi_{0^+_1} \rangle^2  /  \langle \Psi_{1^+_f}  \| \hat{ \mathcal{J}}_{1b} \| \Psi_{0^+_1} \rangle^2 $, for each individual transition with index $f$ and excitation energy $E_f$. The couplings are adopted as $c_3 = -3.2$, $c_4 = 5.4$ and $c_D = -2.0$ in the two-body currents. } 
\end{center}
\end{figure*}

We stress that the inclusion of 4-qp configurations  has significant physical implication for the excitation regions $E_f \ge 3.0$ in the present study, and generally in all cases wherever highly excited states are involved. In the recent example, the same PSM calculation showed that the  inclusion of higher-order qp-states leads to a broad fragmentation in GT strengths in $^{59}$Co, thus successfully reproducing the  charge-exchange data extending to 10 MeV \cite{BSGao_PRC_2025}. It was also shown that to provide a correct description for nuclear level densities that increase {\it exponentially} with energy, inclusion of higher-order qp configurations in the model space is essential \cite{JQWang_PRC_2023}. In the present calculation, states above 3 MeV of excitation roughly correspond to 2-qp configurations across major shells, or to 4-qp configurations correspond to a simultaneous breaking of one neutron and one proton pair. Therefore, the correlations between different major shells as well as among many higher-order qp states are taken into account in the present calculation, which are the decisive  source of suppressing GT strength (see blue curve in Fig. \ref{fig:sum_BGT}).

Finally, we discuss the quenching effect caused by the TBC. In Fig. \ref{fig:sum_BGT}, the colored bands show calculated results considering {\it both} the large configuration space and the OBC plus TBC. These colored bands correspond to the results for different coupling constants suggested in the literature, namely $(c_3, c_4) =$ $(-3.4, 3.4)$ \cite{coupling_NPA_2005}, $(-4.78, 3.96)$ \cite{coupling_PRC_2003}, and $(-3.2, 5.4)$ \cite{coupling_PRC_2003_R}, all in units of GeV$^{-1}$. The error estimate for the bandwidth, expressed as $\bar c_D \equiv \bar d_1 + 2\bar d_2$, ranges from $-2.0$ to $2.0$. Note that in Fig. \ref{fig:sum_BGT}, the calculated red and blue bands almost overlap. From the calculation, we can conclude that, on average, chiral TBC leads to a $5-15\%$ quenching in GT strength, with quenching caused by the coupling constants $(-3.2, 5.4)$ (purple band) being more significant. For $E_f < 1.0$ MeV, the TBC causes little quenching; however, for $1.5 < E_f < 2.0$ MeV, it causes excessive quenching.

Thus, the quenching due to TBC exhibits a sensitive structure dependence, suggesting that the two sources that cause quenching are not disentangled. To see this more clearly, effects of the TBC are shown in Fig. \ref{fig:quenching} for each individual GT strength. One sees that, while the TBC leads to $5-15\%$ quenching for most transitions (green dots), there exist cases showing more than $100\%$ enhancement (red dots) or more than $20\%$ reduction (blue dots). 


What are the states that experience such  extraordinary effects of the TBC? By analyzing the wave functions, we find that when only the OBC is considered, the large B(GT)'s in Fig. \ref{fig:each_BGT}(b) correspond to the transitions to the $1^+$ final states mainly involving deformed low-$K$ levels, namely, $K^\pi = 1/2^-$ levels originating from neutron $1f_{5/2}, 2p_{1/2}$ orbitals, and $K^\pi = 3/2^-$ levels originating from proton $1f_{5/2}, 2p_{3/2}$ orbitals. It is interesting that the two-body current can induce strong mixing in the $1^+$ final states with deformed high-$K$ levels. For the examples of enhanced ratio shown in Fig. \ref{fig:quenching} (red dots), the corresponding wavefunctions show large components from deformed high-$j$ orbit of neutron $1g_{9/2}$, $1g_{7/2}$ and proton $1g_{9/2}$ orbitals. On the other hand, for the cases with large quenching in Fig. \ref{fig:quenching} (blue dots), the corresponding wavefunctions show very strong mixing among many deformed levels originating from both low-$K$ and high-$K$ orbitals. Instead of single strong transitions usually seen in low energies of spherical nuclei, GT strength is distributed over many final states of deformed nuclei with strong fragmentations. The early proposed additional GT selection rules for deformed nuclei \cite{Alaga_PR_1955, Alaga_NP_1957_selection, Mottelson_Nilsson_1959_MFSDVS, Sood_PRC_1992} become completely meaningless.

\begin{figure}
\begin{center}
  \includegraphics[width=0.46\textwidth]{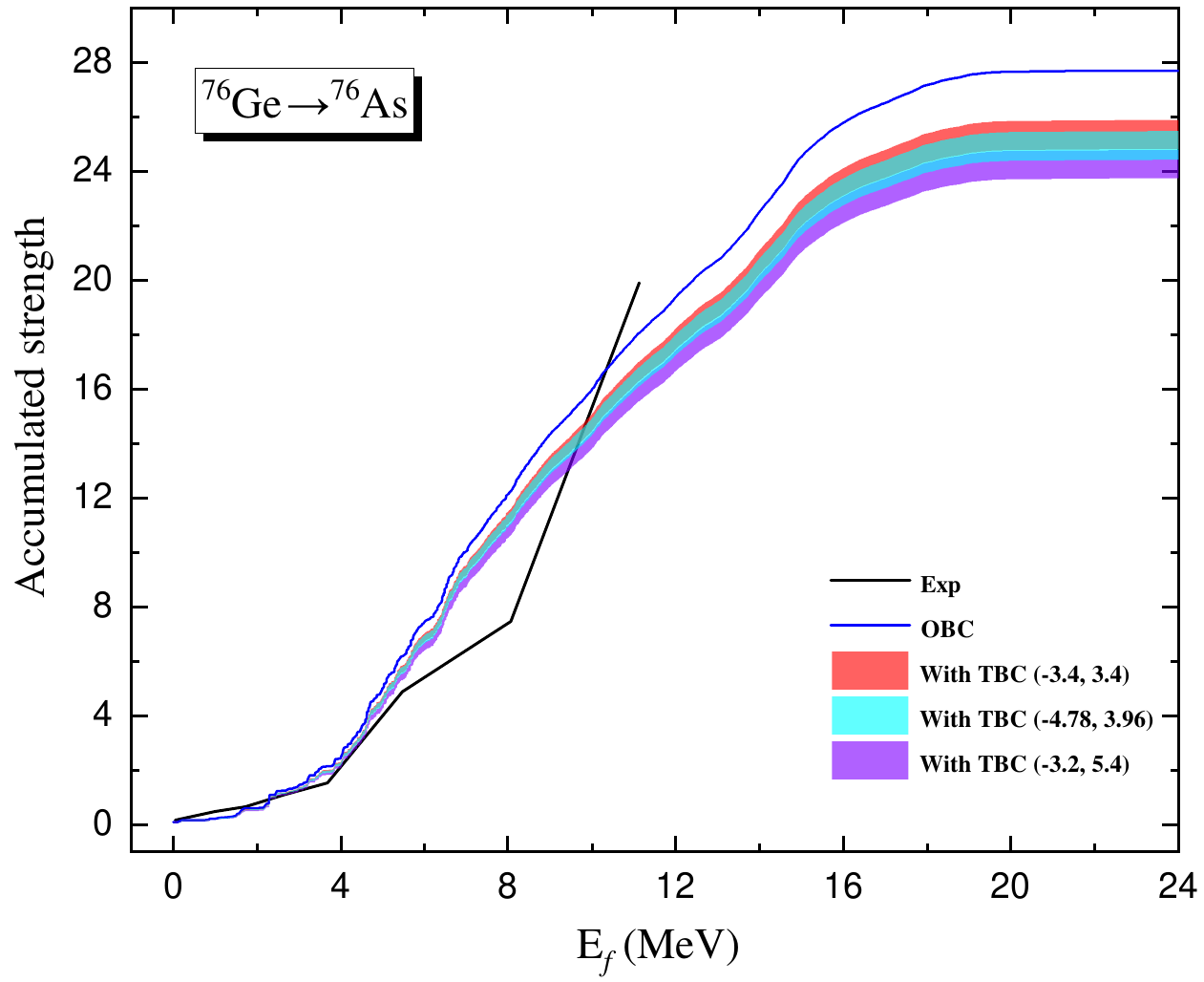}
  \caption{\label{fig:high_lying} (Color online) Effect of the two-body current (TBC) with increasing level density. Cumulative sum of GT transition strengths from the $0_1^+$ ground state of $^{76}$Ge to all $1_f^+$ states of $^{76}$As with $E_f \lesssim 20.0$ MeV, calculated by the PSM with fixed large configuration space and the one-body current (OBC) are compared with the calculations that further considered TBC. The charge-exchange reaction data \cite{Madey_PRC_1989}, without giving error bars, are shown as reference.  } 
\end{center}
\end{figure}


In Fig. \ref{fig:high_lying}, we show cumulative sum of GT transition strengths from the $0^+_1$ ground state of $^{76}$Ge to all the final $1^+_f$ states in  $^{76}$As up to 24 MeV of excitation, where calculations without and with the chiral TBC are compared with each other. The charge-exchange reaction data \cite{Madey_PRC_1989} are shown for reference. In the calculation, we apply an energy cutoff for multi-qp states (see Eq. (\ref{eq.config})) to allow those of $E_{\text{qp}} \lesssim 18.0$ MeV. With this truncation, we obtain about 25,000 $1^+$ levels within $E_f \lesssim 18.0$ MeV, and only a few more $1^+$ levels beyond 18 MeV. 

The results in Fig. \ref{fig:high_lying} indicate a broad enhancement of GT strength visually seen from 5 MeV to 15 MeV, without prominent resonance peaks as typically predicted by QRPA calculations. Our GT strength enhancement is found to consist of thousands of individual transitions within each 1 MeV energy bin. Due to the configuration-space cutoff, the accumulated strength saturates at $E_f > 18$ MeV. As can be seen, the effect of TBC (i.e. the difference between the OBC curve and the colored TBC bands) accumulates at a rate similar to that of the accumulative GT strength. Therefore, we can conclude that even at the highest excited states, TBC still leads to an overall quenching of GT strength by $5-15\%$.  It should be noted that this conclusion is drawn under the inclusion of the exceptional transitions (i.e., the red and blue dots in Fig. \ref{fig:quenching}). This information may be useful for calculations employing the full set of levels (such as nuclear matrix elements in double $\beta$ decays).

In summary, based on a novel shell-model approach that accommodates a large configuration space while
taking the chiral two-body current into account exactly, we have studied several structural effects that had not been thoroughly addressed in conjunction with the quenching problem. The applied code is generally written for allowed GT transition calculations for arbitrarily heavy and deformed nuclei, and for low-lying and highly excited states within and beyond the $Q$ value. The last aspect is important for understanding experimental data of charge-exchange reactions and for theoretical calculations of nuclear matrix elements in double $\beta$ decay.

In the analysis, we have examined, step by step, several structural effects including nuclear deformation, configuration space truncation, and the chiral TBC. For the chosen example $^{76}$Ge, one of the neutrinoless double $\beta$ decay candidates, our calculations show that the choice of the mean-field through nuclear deformation influences significantly the entire GT spectrum. With given deformation, delicate GT strength redistribution occurs between low- and high-excitation regions. With the enlarged multi-qp  configuration space that induces mixing within exponentially increasing number of states as function of excitation (see Fig. \ref{fig:each_BGT}(c)), the overall GT strengths are suppressed considerably while much more newborn states with tiny strengths appear at higher excitations (see Fig. \ref{fig:each_BGT}(b)). Effectively, low-energy GT strengths are quenched and the quenched amount moves to higher excitations. 
Our study thus reveals that the quenching is not a local phenomenon, but a global one involving all low-energy states and highly excited states that satisfy the GT selection rules. Fitting data locally, whether by applying quenching factors or operator renormalization, may thus risk neglecting a consistent description of $\beta$ decay in other excited regions of the same system.


We have found that the effect of TBC depends sensitively on nuclear structure, suggesting that the two sources that cause quenching can not be separated. In our example, TBC leads to little quenching for $E_f < 1.0$ MeV, but for $1.5 < E_f < 2.0$ MeV, it causes excessive quenching. In Fig. \ref{fig:quenching}, we have seen many levels below $E_f < 5$ MeV exhibit extremely high sensitivity to TBC. In the highly excited states, the TBC-sensitivity of individual states is eventually washed out, validating our general conclusion that the chiral two-body current contributes a modest $5–15\%$ reduction of GT strength, depending however on the TBC coupling constant.  

\begin{acknowledgments}
  We thank B.-L. Wang for checking the two-body matrix elements of the chiral two-body currents operators, and J. M. Yao for valuable discussions. This work is supported by the National Natural Science Foundation of China (Grants No. 12275225, No. 12235003). 
\end{acknowledgments}

\bibliography{magnetic_weak_process} 

\end{document}